\documentstyle[12pt,worldsci]{article}
\input epsf.tex
\def\ltap{\ \raise.3ex\hbox{$<$\kern-.75em\lower1ex\hbox{$\sim$}}\ }
\def\gtap{\ \raise.3ex\hbox{$>$\kern-.75em\lower1ex\hbox{$\sim$}}\ }

\def\be{\begin{equation}}
\def\ee{\end{equation}}

\begin{document}
\hfill\hbox{\vbox{\hbox{UW/PT-97/08}\hbox{hep-ph/9704302}}}
\title{CP Violation and FCNC in the Third Family from Effective
  Supersymmetry}\author{Ann E. Nelson\\
{\em Department of Physics 1560, University of Washington,
  Seattle, WA 98195-1560, USA}
}  
\maketitle
\setlength{\baselineskip}{2.8ex}
\abstract{I discuss a ``more minimal'' modification of the minimal
  supersymmetric standard model, in which supersymmetry breaking is
  connected with the physics of flavor. 
  Flavor Changing Neutral Currents (FCNC) for the first two families are suppressed, and for the third
   family may be of comparable size to the FCNC in the Standard
   Model.}
\vskip .2truein

Modifications of the Minimal Standard Model are haunted by a
fundamental dichotomy\cite{flavor}: namely that in most extensions, Flavor Changing Neutral Current
(FCNC) constraints are naturally satisfied only if the new physics
scale is above 10--1000 TeV, whereas natural electroweak symmetry
breaking requires new physics below $\sim$ 1 TeV. In table~1 I
contrast the virtues and omissions of the Standard Model with two popular extensions: the Minimal
Supersymmetric Standard Model (MSSM) with soft supersymmetry breaking,
and  Technicolor. On the left I have listed various experimental
observations that a good theory should explain. It is clear 
that all these models have many
shortcomings, however each of them explains
at least one experimental fact in a way that is so beautiful 
it is hard to believe
nature would not make use of it. Now I would like to discuss what
features of each of these models makes them so successful, and how one
might put them together in a single theory\cite{CKN}.

\begin{figure}[t]
\centerline{\epsfxsize=3 in \epsfbox{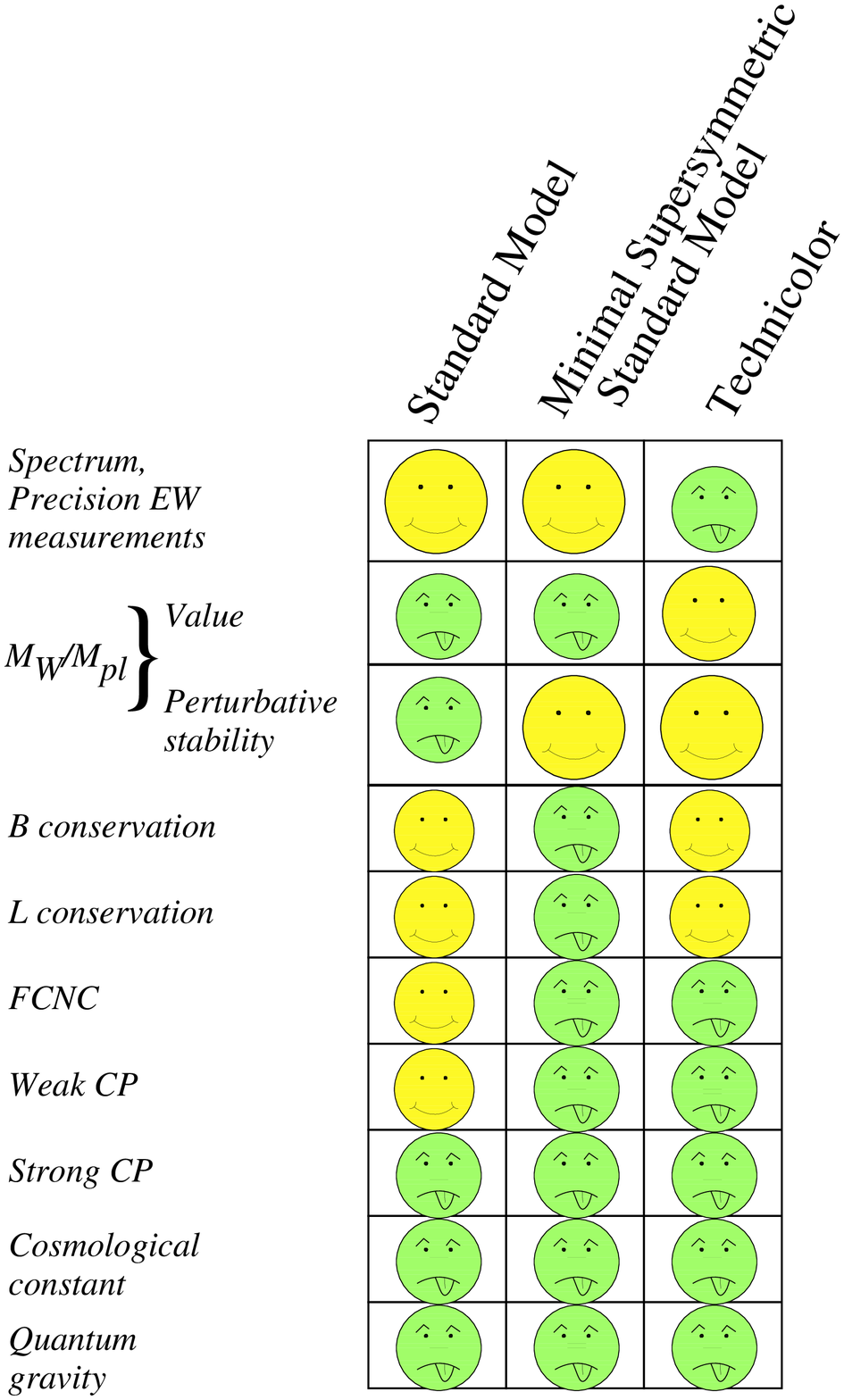}}
\vskip-.2in
\noindent {\bf Table 1}
\vskip .3in
\end{figure}

\section{The observed spectrum and couplings} Both the Standard Model
and the MSSM are very succesful at accomodating the observed fermion
masses, mixings, and couplings to the weak gauge bosons, and the key
to this success is a fundamental Higgs scalar. 

\section{Naturally small $M_W/M_{pl}$}  Here we should borrow
solutions from both the MSSM and from technicolor. Weak scale supersymmetry
naturally
stabilizes the value of the weak scale against perturbative quantum
corrections\cite{pert}, and Dynamical Supersymmetry
Breaking\cite{witten}
 (DSB) can explain why this
scale is so far below the Planck scale, in a manner analogous to the
dynamical electroweak symmetry breaking of Technicolor\cite{WS}.

\section{Baryon Conservation, Lepton Conservation, FCNC, and weak CP
  violation}
Here the most compelling solution is found in the Standard
Model--gauge invariance automatically requires all dangerous operators to be of
dimension higher than 4 and so irrelevant. Baryon and Lepton number
conservation arise as ``accidental'' symmetries, and FCNC are
suppressed by approximate global flavor symmetries which are also
accidental. Unfortunately this is
not at all true in the MSSM, where global symmetries must be imposed
by hand.

\section{The Strong CP Problem, the Cosmological Constant Problem, and
  Quantum Gravity}
None of these issues are addressed by any models for physics at the weak
scale, and it is likely that only a 
theory which is  valid for experimentally inaccessible distance scales
will explain them.

\section{Effective Supersymmetry}
Our conclusion from the above critique is that we should
include both
elementary scalars and supersymmetry in our theory of electroweak
symmetry breaking. In addition, {\it a new  gauge interaction}
is desirable
for two reasons:
\begin{enumerate}
\item to dynamically break supersymmetry
\item to remove the possibility of low dimension B and L violating
  operators by ensuring that any dangerous gauge invariant operators
  are {\it automatically} sufficiently irrelevant.
\end{enumerate}

It is most economical if the same gauge interaction does both jobs, so
the DSB gauge interaction also couples directly to
at least some ordinary fields.   Such  a gauge interaction can give a large
(hopefully positive) supersymmetry breaking mass to the scalar
components of the superfields it couples to. 
If the scalar masses are very large, $\sim  20$ TeV, for the first
two generations of squarks and sleptons, 
the FCNC problems of supersymmetry are alleviated via a new mechanism,
{\it squark and slepton  decoupling} \cite{CKN,decouple}.  

The top squark and left handed bottom squark couple strongly to the
Higgs and natural electroweak symmetry breaking requires that these be
lighter than 1 TeV. Therefore they should not carry any new
supersymmetry breaking gauge interactions. Having these sparticles be
light does not cause any (so far) problematic FCNC. By treating the
generations differently, the dichotomy between the high scale necessary for
FCNC suppression and the lower scale determined by natural electroweak
symmetry breaking is removed.

Note that all dangerous dimension 4 and 5 baryon violating operators
of the MSSM
involve at least one superfield from the first two generations, so the
proton stability could be guaranteed by new gauge symmetries for the first
two generations alone. 

Automatic suppression of dimension 3 and 4 lepton number
violating interactions requires that the down-type Higgs and the
sleptons be distinguished by different gauge interactions as well.
One possibility is for the down type Higgs to carry the new
supersymmetry breaking gauge interaction, which would make
the scalar very heavy and lead to naturally large $\tan\beta$. Another
possibility
is for all generations of left handed leptons to carry the new
interaction and so all the left handed sleptons would be heavy.

Such a new interaction could also suppress the couplings of the
first two generations to the Higgs, explaining the generational mass hierarchy.  
Thus, at least in principle, a new gauge interaction carried by the
first two generations of quarks and leptons could solve a remarable
number of problems. It could produce  the gauge hierarchy via DSB, 
while suppressing FCNC, SUSY contributions to particle
electric dipole moments, and  baryon and lepton number violation. It
could even  explain
the fermion mass hierarchy. Unfortunately no completly explicit and
successful examples exist, but several recently constructed models
come close \cite{CKN,anom}.   

Of course we have to assume that  new gauge symmetries for ordinary
particles
are
spontaneously broken or confined above some scale $\Lambda$. The scale
of heavy squark and slepton masses will then be of order
$\Lambda_S^2/\Lambda\sim 20$ TeV,
where $\Lambda_S$ is the fundamental supersymmetry breaking
scale. The ordinary gauginos, top squarks and left handed bottom
squark, which only could weakly to the supersymmetry breaking sector,
are significantly lighter, below 1 TeV. Other scalars of the
third generation may be either light or heavy. 

\begin{figure}[t]
\centerline{\epsfxsize=4 in \epsfbox{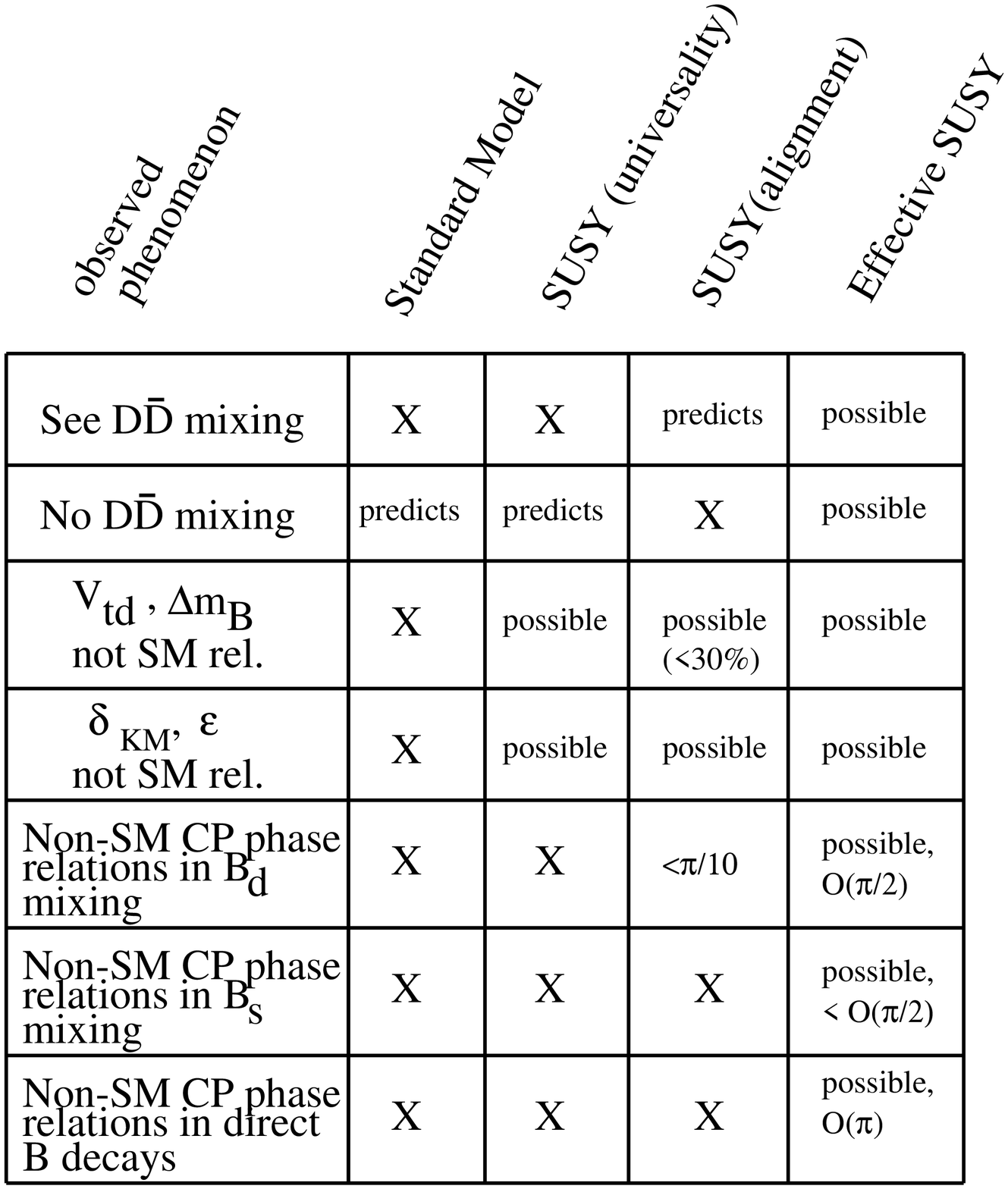}}
\vskip-.2in
\noindent {\bf Table 2}
\vskip .3in
\end{figure}

The presence of a light left handed bottom squark will enhance the
SUSY  box diagram contribution  to $B-\bar B$ mixing. In addition there is no
reason to expect the SUSY phase in this contribution to be the
same as that of the standard model diagram. Similarly, there can be
significant SUSY penguins, with new CP vioalting phases,  for the
third generation. Thus the B factory signals for Effective
Supersymmetry can be quite different from those in the MSSM with universal
squark masses, or with squark masses aligned with the down quark
masses\cite{bphys}.
In Table 2 I summarize the potentially observable differences between
the different SUSY models. Clearly the framework I have described is
not sufficiently predictive to be ruled out by experiments at low
energies. However I am optimistic that  hints of physics beyond the
Standard Model will be found in such experiments.  A decisive test 
will have to wait for the next
generation of collider experiments. Discovery of a squark or slepton
from the first two generations rules out Effective Supersymmetry,
however we expect the Higgs, top squarks,
left handed bottom squark, Higgsinos and gauginos to be found.

\section*{Acknowledgements}
I would like to thank the organizers for inviting me to this most
interesting conference, and David Kaplan for the use of the tables
done on his NEXT. This work was supported in part by the DOE under
grant no. \#DE-FG03-96ER40956.

\end{document}